\documentclass[sn-mathphys]{sn-jnl}

\jyear{2021}%

\theoremstyle{thmstyleone}%

\theoremstyle{thmstyletwo}%

\theoremstyle{thmstylethree}%

\usepackage{lscape}
\begin{document}

\title[Turbulence over perforated plates]{Permeability and turbulence over\\ perforated plates}

\author*{\fnm{Haris} \sur{Shahzad}}\email{h.shahzad@tudelft.nl}

\author{\fnm{Stefan} \sur{Hickel}}\email{s.hickel@tudelft.nl}

\author{\fnm{Davide} \sur{Modesti}}\email{d.modesti@tudelft.nl}

\affil{\orgdiv{Aerodynamics Group, Faculty of Aerospace Engineering},\\ \orgname{Delft University of Technology}, \\\orgaddress{\street{Kluyverweg 2}, \city{Delft}, \postcode{2629 HS}, \state{South Holland}, \country{The Netherlands}}}

\abstract{
We perform direct numerical simulations of turbulent flow at friction Reynolds number $Re_\tau \approx 500-2000$ grazing over perforates plates with
moderate viscous-scaled orifice diameter $d^+\approx40$--$160$ and analyse the relation between permeability and added drag. 
Unlike previous studies of turbulent flows over permeable surfaces, we find that
flow inside the orifices is dominated by inertial effects,
and that the relevant permeability is the Forchheimer and not the Darcy one.
We find evidence of a fully rough regime where the relevant length scale is the inverse of the Forchheimer coefficient, 
which can be regarded as the resistance experienced by the wall-normal flow.
Moreover, we show that, for low porosities, the Forchheimer coefficient can be estimated
with good accuracy using a simple analytical relation.
}

\keywords{wall turbulence, direct numerical simulation, perforated plates, permeable walls, Forchheimer permeability}

\maketitle

\section{Introduction}\label{sec1}

Turbulent flows grazing over permeable surfaces are common in engineering.
Perforated plates, in particular, are used for flow conditioning~\citep{laws1995further}, enhancing heat transfer in heat exchangers \citep{kutscher1994heat}, flame control in combustion chambers~\citep{wei2017277},
aircraft trailing edge noise abatement~\citep{rubio2019mechanisms} and acoustic liners in aircraft engines~\citep{casalino_18,shur_21}. 
Many of these applications feature turbulent grazing boundary layers over perforated plates, which result in higher drag than the baseline smooth wall.
However, the drag increase is often accepted as a mandatory compromise to effectively control some other flow property, such as sound or heat transfer.

Perforated plates are substantially different from other porous surfaces, such as metal foams, ceramic filters and gravel, for example, because
they are characterized by relatively larger pores with respect to the boundary layer thickness of the grazing flow, $d/\delta\approx\mathcal{O}(0.1)$~\cite{avallone_lattice-boltzmann_2019}. 
Pores with large diameter have the potential to substantially alter the flow physics compared to 
canonical porous surfaces which are characterized by large porosity (i.e. open-area ratio), $\sigma>0.8$, but very small pore diameters, 
$d/\delta<0.01$ and $d^+=d/\delta_v < 20$ \cite{efstathiou_18,manes_turbulent_2011-1}, where $\delta_v=\nu/u_\tau$ the viscous length scale, 
$u_\tau=\sqrt{\tau_w/\rho}$ the friction velocity, $\tau_w$ the drag per plane area and $\rho$ and $\nu$ are the fluid density and kinematic viscosity, respectively.
These small pore sizes allows us to accurately model this type of surfaces using Darcy models,

\begin{equation}
	-\nabla P = \dfrac{\mu}{K_{ij}} U_c, \label{eq:darcy}
\end{equation}

where $\nabla P$ is the pressure gradient across the permeable layer, $K_{ij}$ is the permeability tensor, $U_c$ is a characteristic velocity, and $\mu$ is the dynamic viscosity of the fluid. 
Darcy permeability has the physical dimensions of an area, and it represents the ease with which flow passes through a porous surface.
The Darcy equation Equation~\eqref{eq:darcy} stems from the mean momentum balance of the Navier--Stokes equations, and it is usually considered an accurate model of canonical porous surfaces, at least
when the Reynolds number based on the pore diameter is small enough that the underlying Stokes approximation remains valid.

Several authors have used Darcy's boundary conditions to model the turbulent flow over porous substrates~\cite{rosti_turbulent_2018,li_turbulent_2020}
and reported accurate results as compared to pore-resolved simulations~\citep{kuwata_direct_2017}. 
With the exception of some particular configurations~\cite{gomez-de-segura_turbulent_2019}, 
porous surfaces tend to increase drag, similar to surface topography.
Manes \textit{et al.}  ~\cite{manes_turbulence_2009,manes_turbulent_2011} discussed the similarities and differences between canonical porous surfaces and roughness
and concluded that porous surfaces interact differently with the grazing flow as compared to rough surfaces. Breugem \textit{et al.} \cite{breugem_influence_2006}, for instance, report that porous surfaces do not exhibit a fully rough regime in which the skin-friction coefficient approached a constant value with increasing Reynolds number.

The added drag provided by rough surfaces is characterized by the roughness Reynolds number $k^+=k/\delta_v$, where $k$ is the roughness height.
Hence, $k$ is usually regarded as the surface length scale to be compared with the viscous length scale, $\delta_v$ to determine whether the flow is strongly affected by the surface roughness.
For porous surfaces, two types of length scales have generally been considered, namely the pore size $d$ and the square root of the permeabilities $\sqrt{K_{ij}}$, but several
authors have shown that drag depends on the dominant viscous-scaled permeability component, or a combination of $\sqrt{K_{ij}}^+=\sqrt{K_{ij}}/\delta_v$ \cite{gomez-de-segura_turbulent_2019,rosti_turbulent_2018,breugem_influence_2006}.

\begin{figure}
    \centering
        \includegraphics[scale = 0.85] {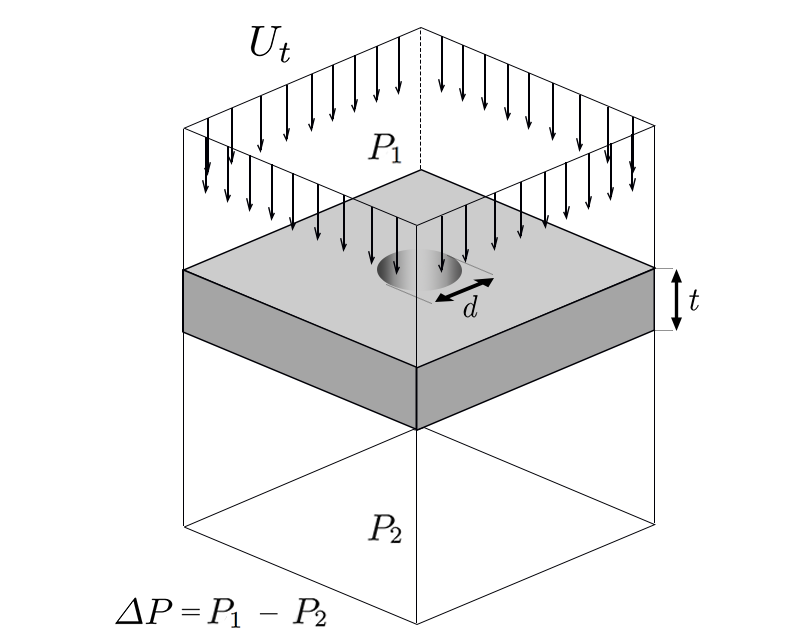} 
        \includegraphics[scale = 0.8] {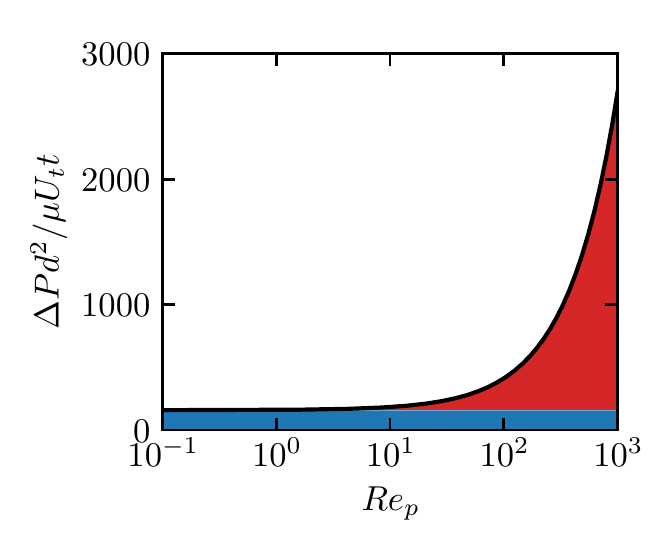} 
	\caption{(\textit{a}) Sketch of the flow normal to a perforated plate with diameter $d$ and thickness $t$. $U_t$ is the superficial velocity and $\Delta P = P_1 - P_2$ the pressure drop through the plate. (\textit{b}) Darcy (blue) and Forchheimer (red) contributions to the pressure drop from Equation~\eqref{eq:pressure_drop} as a function of the pore Reynolds number $Re_p$ 
	for $\sigma = 0.322$ and $t/d = 1$.}
    \label{fig:baeconfig}
\end{figure}

Equation~\eqref{eq:darcy} has been proven to be accurate for many canonical porous surfaces, and it is applicable within the limit of Stokes flow, namely
for small values of the pore Reynolds number $Re_p = \rho d U_p/\mu$, where $U_p$ is the velocity inside the pore.
However, deviations from Darcy's law for increasing $Re_p$ are well documented in the literature \cite{tanner_flowpressure_2019, bae_numerical_2016, lee_empirical_2003}, and have been associated with nonlinear effects that arise at high pore Reynolds numbers.
$Re_p$ is higher for perforated plates than for other porous surfaces and therefore Equation~\eqref{eq:darcy} is replaced by, 

\begin{equation}
	\dfrac{\Delta P}{t} \, \dfrac{d^2}{\mu U_t} = \dfrac{d^2}{K_y} + \sigma \alpha_y d Re_p,
    \label{eq:pressure_drop}
\end{equation}

where $U_t = \sigma U_p$ is the superficial velocity (Figure~\ref{fig:baeconfig}\textit{a}), $t$ is the thickness of the plate and $K_y$ and $\alpha_y$ are
the permeability and the Forchheimer coefficient in the direction normal to the plate, respectively ~\citep{lee_empirical_2003,bae_numerical_2016}.

Figure~\ref{fig:baeconfig}(\text{b}) shows the contribution of the Darcy and Forchheimer terms to the normalized pressure drop as a function of the pore Reynolds number. 
It is evident that nonlinear effects start already at low values of $Re_p\approx5$ and become dominant at $Re_p\approx 50$.

At sufficiently high pore Reynolds number, Darcy drag can be assumed negligible and the entirety of the pressure drag is due to the nonlinear term. 
The pressure drop characteristics of perforated plates at high Reynolds numbers $Re_p \geq \mathcal{O} (10^2)$ 
have been studied extensively both numerically~\cite{tanner_flowpressure_2019} 
and experimentally~\cite{idelchik1986handbook, kast2010l1, malavasi2012pressure, miller1978internal, holt2011cavitation};
however Equation~\eqref{eq:pressure_drop} for the normal flow has never been associated to the case of grazing boundary layer over porous surfaces,
for which Darcy's law has always been used, to our knowledge.

In this study, we aim at clarifying the errors that potentially result from using Equation~\eqref{eq:darcy} for grazing turbulent boundary layers over perforated plates,
and, in particular, from using the square root of the Darcy permeability as a relevant length scale.
First, we perform simulations of laminar flow through perforated plates to compute the Forchheimer permeability coefficients and compare the 
results with experimental and numerical data, and with popular engineering approximations.
Second, we carry out direct numerical simulation (DNS) of turbulent channel flow grazing over perforated walls and discuss the relevance of
the Forchheimer coefficient for the drag of this flow.

\section{Flow through a Perforated Plate}
\label{sec:baecompar}

In order to calculate the Darcy and Forchheimer coefficients, we perform simulations of laminar flow through
a perforated plate using the setup sketched in Figure~\ref{fig:baeconfig}.
We solve the incompressible Navier--Stokes equations,
and fix the superficial velocity at the inflow and the pressure at the outflow.
Neumann boundary conditions are used for the outflow velocity and inflow pressure,
no-slip boundary condition is used at the surface of the perforated plate, and symmetry boundary conditions are used at the lateral boundaries.

Simulations discussed in this section are performed with the pimpleFoam solver, which is part of the open-source library OpenFOAM$^\text{\textregistered}$~\citep{weller_98}. 
A forward Euler time step scheme with a maximum CFL number of 0.7 is used and simulations are run until a steady-state solution is reached (residual $< 10^{-9}$).
The inflow and outflow boundaries are at least 40 orifice diameters away from the perforated plate. We have verified that the final solution
is independent of the domain size.
Approximately 10M cells are used with a minimum mesh size of $\approx0.001d$ in the proximity of the plate orifice. 
We have performed a grid resolution study to ensure that the presented results are fully converged.

We consider 9 plate geometries with different porosity and thickness-to-diameter ratio $t/d$, which are summarised in Table~\ref{tab:permcases}.
Six geometries are designed to match the parameters
of Bae and Kim \cite{bae_numerical_2016} ($B_{r1}-B_{r3}$), and Tanner \textit{et al.} \cite{tanner_flowpressure_2019} ($T_{r1}-T_{r3}$), whereas
($L_{g1}-L_{g3}$) are novel geometries.
Permeability is considered independent of the spacing of the holes~\cite{bae_numerical_2016,malavasi2012pressure,tanner_flowpressure_2019}, therefore we simulate plates with a single orifice, and change the porosity by changing the orifice diameter. 
The pressure drop $\Delta P$ is evaluated as the difference between the inlet and outlet pressure, cf the schematic in Figure~\ref{fig:baeconfig}(\textit{a}).
For each of the 9 plate geometries , we perform simulations at different $Re_p$ and use Equation~\eqref{eq:pressure_drop} to compute 
the Darcy permeability and the Forchheimer coefficient. 

\begin{landscape}
\begin{table}
\centering
\begin{tabular}{cccccccccc}
\hline
         &      $\sigma$    &  $t/d$     & \multicolumn{7}{c}{$\alpha_y \,d$}                                                                \\
\hline
	 &  &  & Kast \textit{et al.} \cite{kast2010l1} & Idelchik \cite{idelchik1986handbook} & Malvasi \textit{et al.} \cite{malavasi2012pressure} & Miller \cite{miller1978internal} & Holt \textit{et al.} \cite{holt2011cavitation}  & Bae and Kim \cite{bae_numerical_2016} & Present  \\
\hline
$B_{r1}$ & 0.2      & 2     & 12.3 & 6.59 & 12.5 & 6.55 & 4.82  & 7.5  & 7.68                                               \\
$B_{r2}$ & 0.3      & 2     & 4.69 & 2.36 & 4.62 & 2.35 & 1.75  & 2.91 & 4.67                                               \\
$B_{r3}$ & 0.4      & 2     & 2.25 & 1.05 & 2.07 & 1.02 & 0.791 & 1.41 & 2.61                                               \\
$T_{r1}$ & 0.2      & 0.25  & 98.0 & 94.9 & 100  & 247  & 89.9  & 60.0 & 70.2                                               \\
$T_{r2}$ & 0.4      & 0.25  & 18.0 & 15.2 & 16.6 & 38.5 & 13.4  & 11.2 & 13.17                                              \\
$T_{r3}$ & 0.6      & 0.25  & 5.55 & 3.69 & 3.91 & 8.28 & 3.69  & 3.33 & 4.43                                               \\
$L_{g1}$ & 0.0357   & 1     & 960  & 643  & 1005 & 644  & 604   & 567  & 763                                                \\
$L_{g2}$ & 0.143    & 1     & 51.9 & 33.1 & 53.8 & 33.6 & 25.0  & 31.4 & 46.2                                               \\
$L_{g3}$ & 0.322    & 1     & 7.89 & 4.51 & 7.59 & 3.95 & 3.00  & 4.91 & 7.87                                               \\
\hline
\end{tabular}
\caption{Forchheimer coefficient for different porosities $\sigma$ and thickness-to-diameter ratio $t/d$. 
	The last column refers to the present dataset, whereas the other columns refer to the values obtained with popular engineering formulas reported in the Appendix~\ref{secA1}.}
\label{tab:permcases}
\end{table}
\end{landscape}

\begin{figure}
    \centering
        \includegraphics[scale = 1] {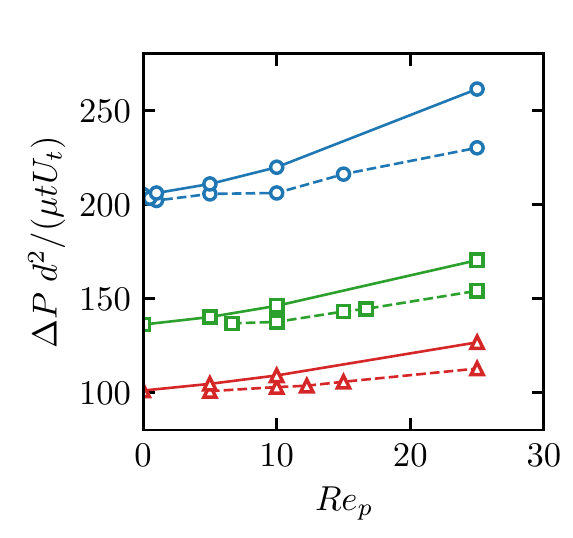}
        \put(-165,145){(\textit{a})}
        \includegraphics[scale = 1] {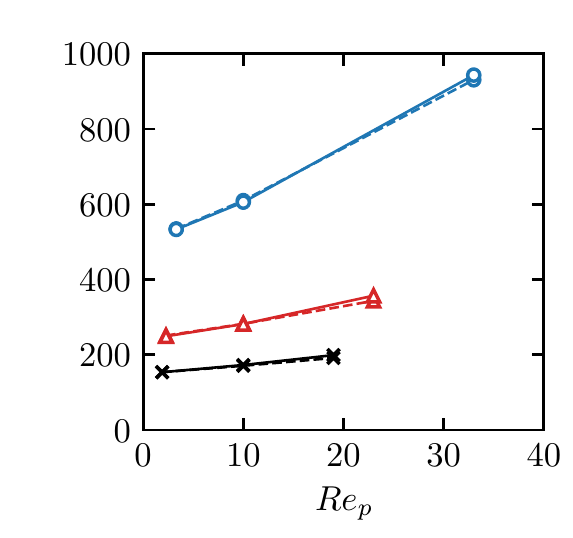}
        \put(-165,145){(\textit{b})} \\
	\caption{Normalised pressure drop as a function of the pore Reynolds number for cases $B_{r1}$-$B_{r3}$ (\textit{a}) and cases $T_{r1}$-$T_{r3}$ (\textit{b}). 
	Solid lines represent pressure drop for the current simulations. Dashed lines represent the pressure drops by Bae and Kim \cite{bae_numerical_2016} and Tanner \textit{et al.} \cite{tanner_flowpressure_2019} in (\textit{a}) and (\textit{b}), 
	respectively. Symbols refer to different porosities: $\sigma = 0.2$ (circles), $\sigma = 0.3$ (squares), $\sigma = 0.4$ (triangles) and $\sigma = 0.5$ (crosses).}
    \label{fig:pressuredrop}
\end{figure}

Figure~\ref{fig:pressuredrop} 2 shows the pressure drop of flow cases $B_{r1}-B_{r3}$(\textit{a}) and $T_{r1}-T_{r3}$(\textit{b})
for our simulations and corresponding data from Bae and Kim ~\cite{bae_numerical_2016} and Tanner \textit{et al.} \cite{tanner_flowpressure_2019}.
We note a disagreement for flow cases $B_{r1}-B_{r3}$ when compared to the data of Bae and Kim \cite{bae_numerical_2016}, which
becomes more evident for increasing pore Reynolds number, with differences up to $20\%$ at the highest $Re_p$.
On the contrary, we observe a very good match for flow cases $T_{r1}-T_{r3}$ with the data of Tanner \textit{et al.} ~\cite{tanner_flowpressure_2019}.
The reasons for the mismatch between the two datasets can be numerous. However, discrepancies of this order of magnitude
are possible at high $Re_p$~\cite{malavasi2012pressure}, and therefore we consider the accuracy acceptable.

As additional validation, we compare the Forchheimer coefficients from the current simulations to several engineering correlations based on experimental data, which are summarized in Appendix~\ref{secA1}.
The values of $\alpha_y$ returned by these correlations are reported in Table \ref{tab:permcases}.
There is a large spread in the Forchheimer coefficient proposed by the different correlations, and differences up to $50$--$60\%$ seem common in the literature.

\begin{figure}
    \centering
        \includegraphics[scale = 1] {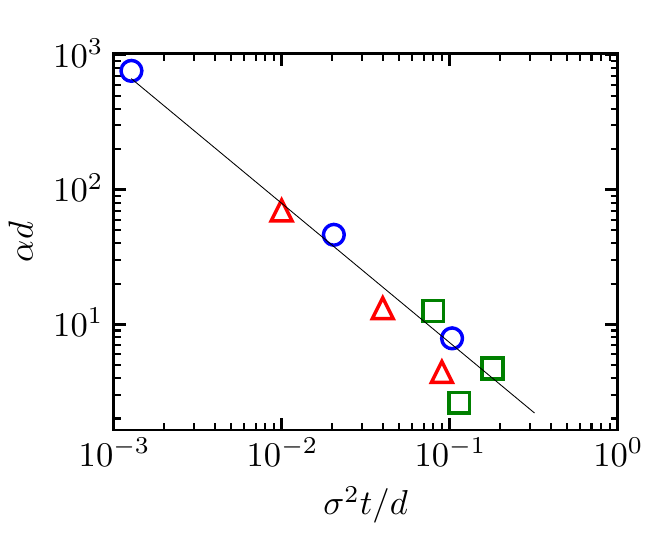}
	\caption{Forchheimer coefficient $\alpha_y\, d$ as a function of $\sigma^2 t / d$ for $L_{g1}$-$L_{g3}$ (circles), $L_{r1}$-$L_{r3}$ (squares) and $T_{r1}$-$T_{r3}$ (triangles).} 
    \label{fig:approxrelation}
\end{figure}

This large uncertainty of the Forchheimer coefficient can be traced back to the weak dependence of $\alpha_y$ on $Re_p$, which has been reported by several studies~\cite{tanner_flowpressure_2019}.
Most of these empirical correlations
are based on data at high pore Reynolds numbers in the attempt to minimize the dependence on $Re_p$. However, this is often not enough
because the dependence of $\alpha_y$ on $Re_p$ can be more or less significant depending on the thickness-to-pore ratio $t/d$ \cite{tanner_flowpressure_2019},
thus complicating the evaluation of the Forchheimer coefficient.
Perfect agreement with the empirical correlations is therefore not expected.
However, the Forchheimer coefficients we calculated 
lie approximately within the range of the different empirical approximations and within the range of their uncertainty.

Even though these correlations differ from each other, they all suggest the same trend of the Forchheimer coefficient for low values of $\sigma$, namely $\alpha_y \sim 1/(\sigma^2 t)$. For this reason, we report $\alpha_yd$ as a function of $\sigma^2 t/d$ in Figure~\ref{fig:approxrelation}.
The Figure shows a visual representation of the Forchheimer coefficient which highlight that the dependence of $\alpha_y$
on the geometry can be condensed in a single independent variable.

\section{Turbulent flow over perforated plates}

In this section, we present DNS results of turbulent grazing flow over perforated plates for different porosities and Reynolds numbers.
Even though perforated plates are an elementary porous surface in terms of geometry, several configurations are in principle possible.
Here, we consider geometries that resemble the acoustic liners used within aircraft engines, which consist of a perforated facesheet and a solid backplate with a honeycomb in between.
Acoustic liners have an orifice diameter of about $d/\delta\approx0.1$ and a honeycomb depth $h/\delta = 4$, 
a porosity in the range $\sigma=0.05-0.3$, and a plate thickness of $t/d\approx1$.

\subsection{Methodology}

\begin{table}[t]
\centering
\resizebox{1\linewidth}{!}{
\begin{tabular}{cccccccccccc}
       & $Re_b$ & $Re_\tau$ & $d^+$ & $\sigma$ & $\sqrt{K_y}^+$ & $1/\alpha_y^+$ & $\Delta U^+$ & $\Delta x^+$ & $\Delta y^+_\text{min}$ & $\Delta y^+_\text{max}$ & $\Delta z^+$  \\ 
\hline
$S_1$  & 9268   & 506.1     & 0     & 0        & 0              & 0              & -            & 5.1          & 0.80                    & 3.83                    & 5.1           \\
$S_2$  & 21180  & 1048      & 0     & 0        & 0              & 0              & -            & 5.2          & 0.80                    & 4.45                    & 5.2           \\
$S_3$  & 45240  & 2060      & 0     & 0        & 0              & 0              & -            & 5.2          & 0.80                    & 6.67                    & 5.2           \\
$L_1$  & 9139   & 503.5     & 40.3  & 0.0357   & 1.04           & 0.0528         & 0.14         & 1.1          & 0.80                    & 5.81                    & 1.1           \\
$L_2$  & 8794   & 496.4     & 39.7  & 0.142    & 2.06           & 0.859          & 0.56         & 1.0          & 0.80                    & 5.81                    & 1.0           \\
$L_3$  & 8264   & 505.3     & 40.4  & 0.322    & 3.22           & 5.14           & 1.90         & 1.0          & 0.81                    & 5.81                    & 1.0           \\
$L_4$  & 19505  & 1038      & 83.0  & 0.142    & 4.30           & 1.718          & 0.96         & 2.1          & 0.83                    & 6.30                    & 2.1           \\
$L_u4$ & 19505  & 1044      & 83.5  & 0.142    & 4.32           & 1.727          & 0.98         & 5.9          & 0.84                    & 6.10                    & 5.9           \\
$L_5$  & 17810  & 1026      & 82.1  & 0.322    & 6.53           & 10.4           & 2.78         & 2.1          & 0.82                    & 6.29                    & 2.1           \\
$L_6$  & 35470  & 2044      & 164.0 & 0.322    & 13.0           & 20.8           & 4.44         & 4.1          & 0.82                    & 6.70                    & 4.1           \\
\hline
\end{tabular}
}
\caption{DNS dataset comprising smooth, $(S_n)$ and liner $(L_n)$ cases. $\sigma$ is the porosity (open area ratio), $d^+$ is the orifice diameter, $K_y$ is the Darcy permeability, $\alpha_y$ is the Forchheimer coefficient and $\Delta U^+$ is the Hama roughness function. Simulations are performed in computational a box with dimensions $L_x \times L_y \times L_z = 3\delta \times 3\delta \times 1.5\delta$. $\Delta x^+$ and $\Delta z^+$ are the viscous-scaled mesh spacing in the streamwise and spanwise direction. $\Delta y^+_{\text{min}}$ and $\Delta y^+_{\text{max}}$ are the minimum and the maximum mesh spacing in the wall normal direction. Liner cases $(L_1)$--$(L_6)$ have equispaced orifices in the streamwise and spanwise direction. Case $L_{u4}$ has the same porosity, and orifice size of $L_4$ but the holes are not equispaced in the streamwise direction, Figure \ref{fig:geometry_dns}.}
	\label{tab:cases}
\end{table}

\begin{figure}
    \centering
        \includegraphics[scale = 1] {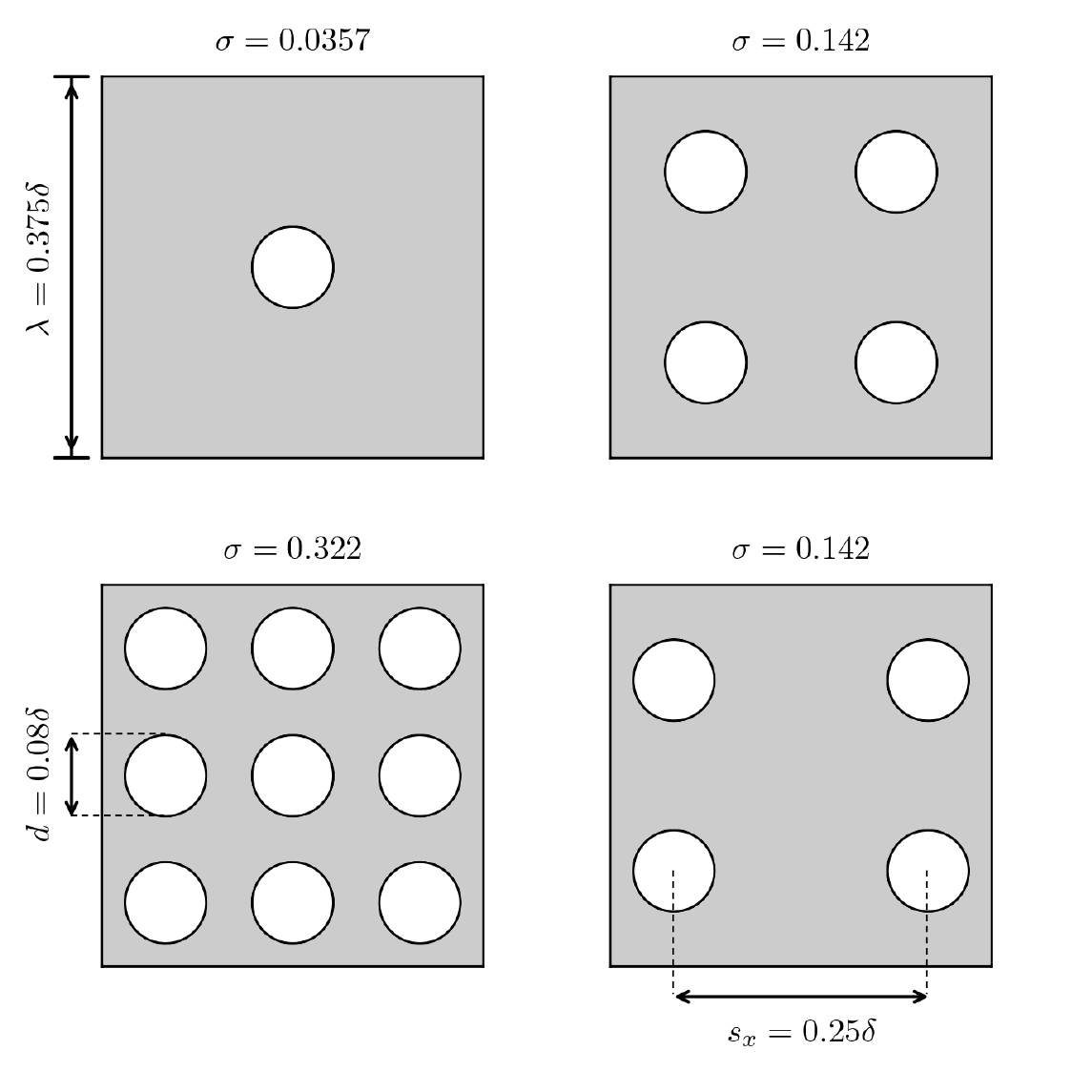}
        \put(-330,320){(\textit{a})}
        \put(-170,320){(\textit{b})}
        \put(-330,160){(\textit{c})}
        \put(-170,160){(\textit{d})}
    \caption{Sketch the holes arrangement in the wall-parallel plane for the DNS cases. 
	Panels (\textit{a}), (\textit{b}) and (\textit{c}) correspond to cases $L_{r1}$, $L_{r2}$ and $L_{r3}$, respectively. 
	Geometries (\textit{a}), (\textit{b}) and (\textit{\textit{c}}) have equispaced holes.
	Geometry (\textit{d}) has the same orifice diameter, plate thickness and porosity as (\textit{b}), but the orifices are not evenly spaced.}
    \label{fig:geometry_dns}
\end{figure}

For the DNS, we solve the compressible Navier--Stokes equations for a perfect gas using the flow solver STREAmS~\cite{bernardini_21}.
The simulations are carried out in a rectangular box of size $L_x \times L_y \times L_z = 3\delta \times 3\delta \times 1.5\delta$, where $\delta$ is the channel half-width. 
The simulations are performed at bulk Mach number, $M_b = u_b/c_w = 0.3$, where $u_b$ is the bulk flow velocity and $c_w$ is the speed of sound at the wall.
At this Mach number, compressibility effects are very small, and the flow can be regarded as representative of incompressible turbulence.
The flow is driven in the streamwise direction by a spatially uniform body force, adjusted every time step to keep a constant bulk velocity $u_b$.

Periodic boundary conditions are applied in the streamwise and spanwise directions and no-slip isothermal boundary conditions are applied at the wall
using a ghost-point immersed boundary method~\citep{vanna_sharp-interface_2020}. 

We choose the liner geometry to match the orifice size of acoustic liners in operating conditions as close as possible. 
The acoustic liner comprises of 64 cavities: an array of $8 \times 4$ in the streamwise and spanwise direction on the upper and lower wall. 
Each cavity has a square cross-section with a side length $\lambda = 0.375 \delta$, 
the orifices have a diameter of $d = 0.08 \delta$, the cavity walls have a thickness of $0.5d$, and the facesheet has a thickness of $d$. 
The cavities have a depth $h = 0.5 \delta$, which is smaller than the one of typical acoustic liners. However, the cavity depth only plays a role
for tuning sound attenuation and not for the aerodynamic drag~\cite{howerton_acoustic_2015}.

We carry out simulations at three friction Reynolds numbers in the range $Re_\tau = \delta/\delta_v \approx 500 - 2000$, 
corresponding to a viscous-scaled diameter of $d^+ \approx 40 - 160$. 
Additionally, we increase the liner porosity between $\sigma= 0.0357$--$0.322$ by varying the number of orifices per cavity between 1 and 9. 
The geometries considered are visualised in Figure \ref{fig:geometry_dns} and the complete list of flow cases is reported in Table~\ref{tab:cases}. 
We also change the spacing between the orifices, flow cases $L_4$ and $L_{u4}$, Figures~\ref{fig:geometry_dns}(\textit{b}),(\textit{d}).
We compare the results of the liner simulations with smooth-wall simulations at approximately matching friction Reynolds numbers. 
Quantities that are non-dimensionalised by $\delta_v$ and $u_{\tau}$ are denoted by the ‘$+$’ superscript. 

\subsection{Added drag and permeability}

As customary for turbulent flows over rough and porous surfaces,
we quantify the added drag with respect to the smooth wall using the viscous-scaled velocity deficit in the logarithmic region $\Delta U^+$, 
also referred to as Hama roughness function~\cite{chung_21}. 
For our geometries, we consider three candidate Reynolds numbers for scaling $\Delta U^+$, namely based on
the orifice diameter $d^+$, based on the square root of the wall-normal permeability $\sqrt{K_y}^+$ and based on the inverse of the Forchheimer coefficient $1/\alpha_y^+$.

\begin{figure}
    \centering
        \includegraphics[scale = 1] {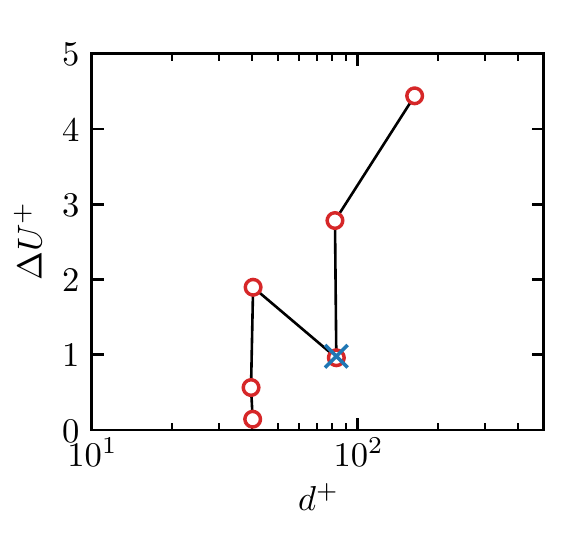}
        \put(-165,145){(\textit{a})}
        \includegraphics[scale = 1] {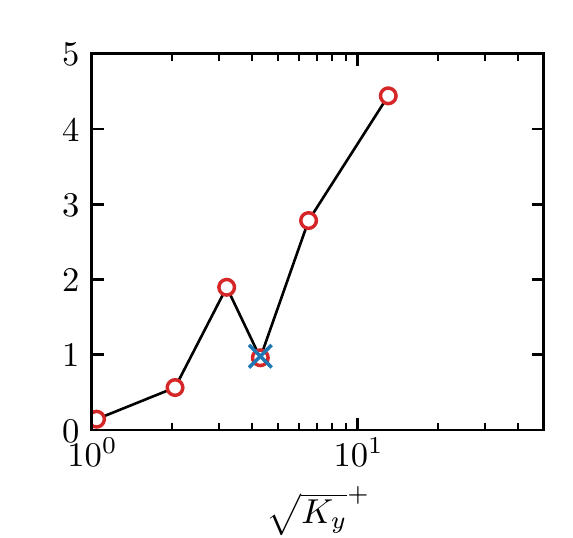}
        \put(-165,145){(\textit{b})} \\
    \caption{$\Delta U^+$ as a function of the viscous-scaled orifice diameter, $d^+$ (\textit{a}) and the Darcy permeability (\textit{b}). 
	Equispaced orifice cases are represented by circles. Non-equispaced orifices case is represented by a cross.}
    \label{fig:deltau_wrong}
\end{figure}

Figure \ref{fig:deltau_wrong}(\textit{a}) shows $\Delta U^+$ as a function of the viscous-scaled diameter. It is clear that $d^+$ alone is not a suitable similarity parameter because increasing the surface porosity for a constant viscous-scaled 
orifice diameter leads to higher $\Delta U^+$. 
For instance, cases $L_2$ and $L_3$ have approximately matching $d^+$, but case $L_3$ exhibits a larger $\Delta U^+$ owing to the higher porosity. 

\begin{figure}
    \centering
        \includegraphics[scale = 1] {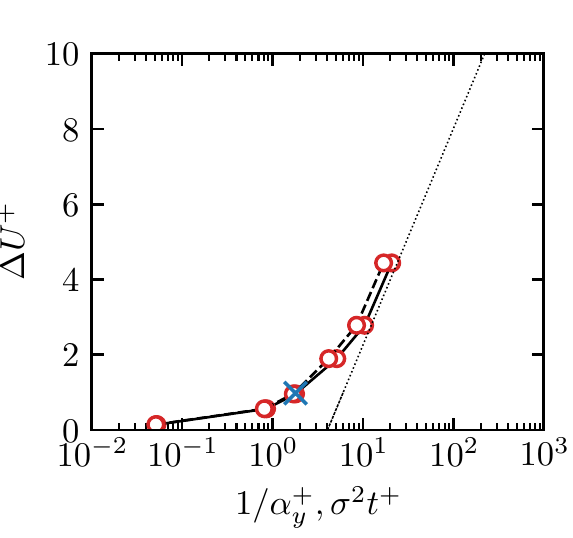}
        \put(-165,145){(\textit{a})}
        \includegraphics[scale = 1] {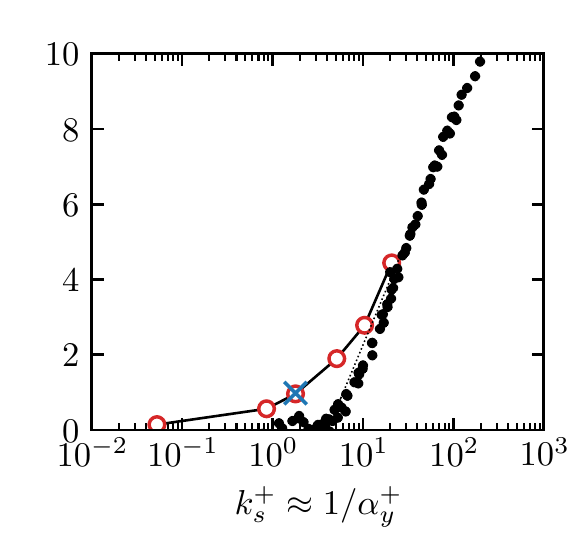}
        \put(-165,145){(\textit{b})} \\
    \caption{$\Delta U^+$ as a function of the inverse of the Forchheimer coefficient, $1/\alpha_y^+$ (solid line) and $\sigma^2 t$ (dashed line) (\textit{a}) and the sandgrain roughness Reynolds number 
	$k_s^+$ (\textit{b}). 
	The dotted line indicates the fully rough asymptote $\kappa^{-1} \text{log}(1/\alpha_y^+)-3.5$. 
	Equispaced orifice cases are represented by circles. Non-equispaced orifices case is represented by a cross.}
    \label{fig:deltau}
\end{figure}

Figure \ref{fig:deltau_wrong} (\textit{b}) shows $\Delta U^+$ as a function of the viscous-scaled wall-normal Darcy permeability. 
The Darcy coefficient is also not suitable for predicting the drag increase as it does not show a consistent monotonic trend. 
Instead, we find that the inverse of the viscous-scaled Forchheimer coefficient $1/\alpha_y^+$ scales very well the effect of the liner, as shown in Figure \ref{fig:deltau} (\textit{a}). 
We clearly see that $1/\alpha_y^+$ is a promising length scale for characterising the additional drag. 
We note that the Hama roughness function tends towards the fully rough regime,
and flow case $L_6$ lies on the lower edge of the asymptote $\Delta U^+\approx \kappa^{-1}  log(1/\alpha_y^+) - 3.5$,
where $\kappa$ is the von K\'arm\'an constant.
Furthermore, Table \ref{tab:cases} shows that case $L_{u4}$ has a nearly identical $\Delta U^+$ to case $L_4$. 
The spacing of the orifices, therefore, has a no effect and the added drag is only a function of $1/\alpha_y^+$,
which provides further evidence that the Forchheimer coefficient is the relevant length scale.
Figure~\ref{fig:deltau}(\textit{b}) shows $\Delta U^+$ of the liner cases as a function of $1/\alpha_y^+$, compared to Nikuradse data 
of sandgrain roughness \cite{nikuradse_stromungsgesetze_1933}, which suggest that acoustic liners behave like sandpaper, as $1/\alpha_y^+\approx k_s$.

Additionally, we test the accuracy of the semi-empirical scaling introduced in Section~\ref{sec:baecompar}, $\alpha_y \approx 1/(\sigma^2 t)$, and 
we plot $\Delta U^+$ as a function of $\sigma^2 t^+$ in Figure~\ref{fig:deltau}(\textit{a}). 
The empirical correlation is very accurate for low values of $\sigma^2 t^+$, whereas minor discrepancies appear
as $\Delta U^+$ approaches the fully rough regime.
This is due to the approximate correlation of the Forchheimer coefficient with $1/(\sigma^2 t)$, as can also be observed  from the formulas
in Appendix~\ref{secA1}. 

\begin{figure}
    \centering
        \includegraphics[scale = 1] {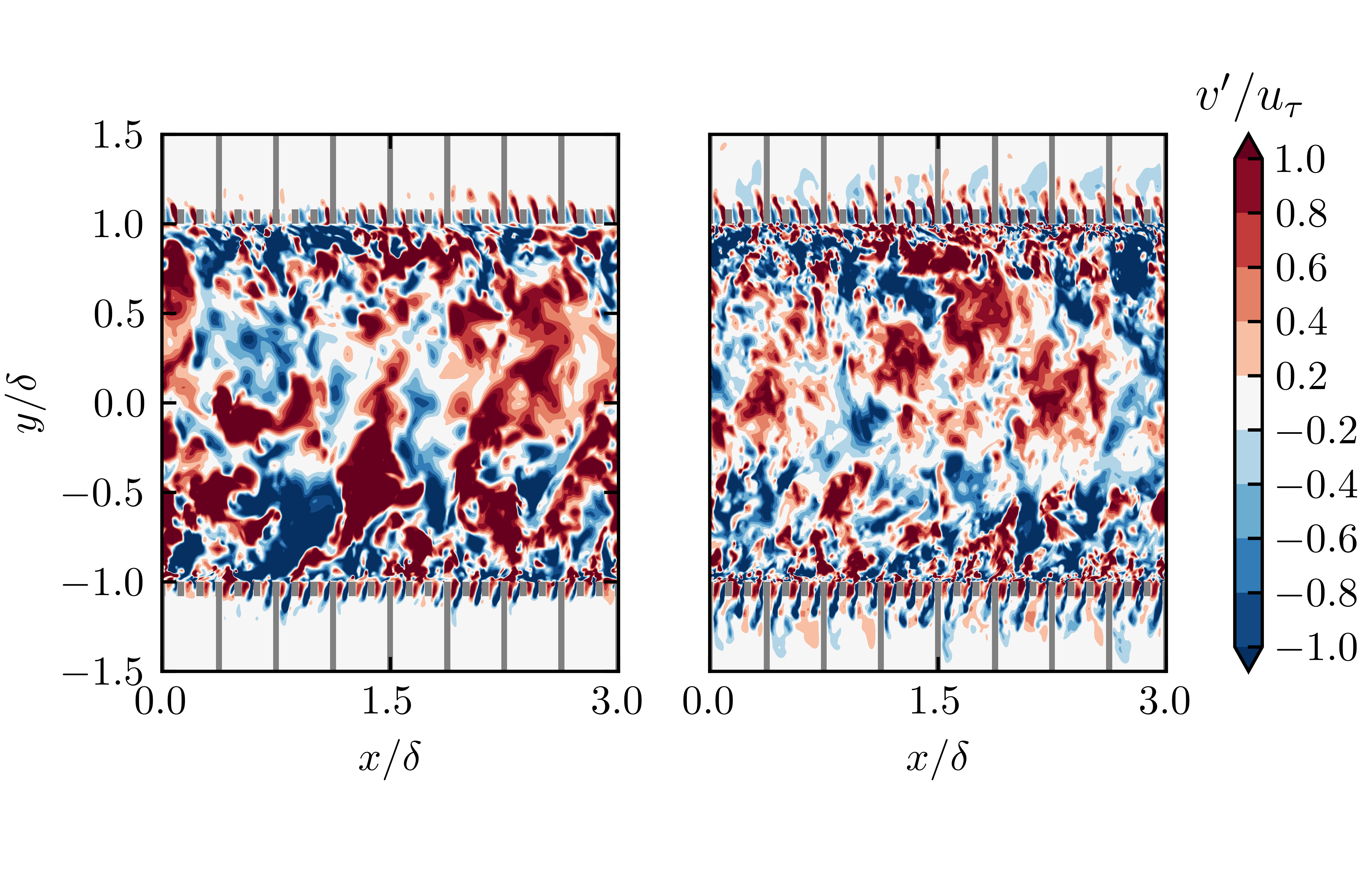}
    \caption{Wall-normal velocity fluctuations in an $x-y$ plane at for flow case $L_4$ (\textit{a}) at $Re_{\tau} \approx 1000$ and flow case $L_5$ (\textit{b}) at $Re_{\tau} \approx 2000$. Grey patches represent solid wall regions.}
    \label{fig:fluc_xy}
\end{figure}

Both the Darcy permeability and the Forchheimer coefficient are suitable candidates to be considered as relevant length scales because
they incorporate the effect of changes in the geometry. However, the Forchheimer coefficient clearly shows superior accuracy for the flow cases under scrutiny.
This can be associated with the relevance of inertial effects inside the orifices, as we qualitatively show in Figure~\ref{fig:fluc_xy}
where we report the instantaneous wall-normal velocity for cases $L_5$ and $L_6$ in a streamwise wall-normal plane.
We observe very high wall-normal velocities inside the orifices forming a jet-like flow from the downstream edge of the orifice into the cavity,
which is particularly evident for flow case $L_6$ as fluid is pushed further inside the cavity.

Using the maximum wall normal velocity fluctuation $v_{rms}$ inside the orifice, we estimate a pore Reynolds number $Re_p \approx 50-500$, depending upon the flow case considered. 
The Forchheimer drag constitutes about 50\% of the total drag at $Re_p \approx 50$ and almost the entirety of the drag at $Re_p \approx 500$, see Figure~\ref{fig:baeconfig}\textit{b}. 
This is further confirmation for the use of the Forchheimer coefficient rather than the Darcy permeability as the relevant length scale for the present flow cases.

To further clarify on the relevance of the nonzero wall-normal velocity on $\Delta U^+$ we recall that the pressure drop through the plate 
can be expressed in the form of friction factor in the wall-normal direction,

\begin{equation}
    f_y = \dfrac{\Delta P}{0.5 \rho U_t^2}. \label{eq:friction}
\end{equation}

In the limit of high Reynolds number, the entirety of the pressure drop can be attributed to the Forchheimer pressure drop
and using Equation \eqref{eq:friction} and Equation \eqref{eq:pressure_drop} it is easy to show that 

\begin{equation}
    \alpha_y = \dfrac{f_y}{2t}.
	\label{eq:fric_fac}
\end{equation}

Hence, $1/\alpha_y$ represents the drag experienced by 
the flow normal to the plate, suggesting that $\Delta U^+$ is intrinsically related to the wall-normal velocity fluctuations. 
This result is consistent with previous studies on rough surfaces that discuss the correlation between drag and wall-normal velocity fluctuations~\cite{orlandi_06},
and it reveals several similarities between roughness and porous surfaces, which have not been reported in the literature so far.

\section{Concluding remarks}

We have analysed the correlation between wall-normal permeability and wall-parallel drag in turbulent flows over perforated plates.
Perforated plates are different from other types of porous surfaces because their porosity does not exceed $\sigma\approx0.3$ in most engineering applications, as higher values
would substantially affect the structural integrity of the plate. Another main difference with respect to other porous surfaces is that the pore Reynolds number
can be large, and in many applications, $Re_p\sim \mathcal{O}(10^2)$ or higher. 
The result is that the Darcy equation does not hold because inertial effects inside the orifice are dominant, and
the ease with which the fluid passes through the plate is better represented by the Forchheimer coefficient than by the Darcy permeability.

Accurate calculation of the Forchheimer coefficient for perforated plates is challenging, and discrepancies up to $50\%$ are common in the literature,
both from numerical and experimental sources.
We calculate the Forchheimer coefficient using numerical simulations, and our results are in good agreement with a subset of the available data and engineering correlations.
Semi-empirical relations for estimating the Forchheimer coefficient often show a complex dependence on the plate geometry. However, we note that in the limit of small porosity
all correlations return the same functional dependence $\alpha_y\sim 1/\sigma^2 t$, which can be used as a first-order approximation.

In order to show the practical relevance of the Forchheimer coefficient in a realistic engineering application,
we carry out direct numerical simulation of turbulent grazing flow over perforated plates, which resemble the acoustic liners used for noise attenuation over aircraft engines.
We show that the inverse of the viscous-scaled Forchheimer coefficient $1/\alpha_y^+$ is the relevant inner Reynolds number for this type of surface, and
the Hama roughness function shows clear evidence of a fully rough regime. Moreover, these perforated plates provide
the same drag as sandgrain roughness with $k_s^+ \approx 1/\alpha_y^+$.
The ability of $1/\alpha_y^+$ to represent the drag of the plate is attributed to the high values of the pore Reynolds number based on the wall-normal velocity fluctuations $Re_p=50$--$500$,
which suggest dominant inertial effects inside the orifice. The high r.m.s wall-normal velocity is immediately noted in the instantaneous flow visualisations.

We believe that this study sheds new light on to the interactions of a turbulent boundary layer flow with porous surfaces. 
We have identified the inverse of Forchheimer coefficient as a highly relevant scaling parameter, and future efforts should be directed towards
an accurate numerical characterization of this length scale, both experimentally and computationally. 
Last but not least, we note that our findings have been verified for a considerably large data set, however, 
this data can cover only a fraction of the cast parameter space.\\ 

\noindent
{\bf Acknowledgments}
We acknowledge PRACE for awarding us access to Piz Daint, at the Swiss National Supercomputing Centre (CSCS), Switzerland.

\section{Declarations}
\subsection{Competing Interests}

No funding was received for conducting this study. All authors certify that they have no affiliations with or involvement in any organization or entity with any financial interest or non-financial interest in the subject matter or materials discussed in this manuscript.

\begin{appendices}

\section{Empirical Correlations for Pressure Drop through Perforated Plates}\label{secA1}

In this Appendix we report popular engineering formulas for estimating the Forchheimer coefficient or the friction factor.

Bae and Kim \cite{bae_numerical_2016} performed numerical simulations of flow through perforated plates and, proposed the following expression for the Forchheimer coefficient:

\begin{equation}
    \alpha_y = \dfrac{3(1-\sigma)}{4\sigma^2 t}. \label{eq:baenondarcy}
\end{equation}

Several experimental studies at high Reynolds number are available which provide semi-empirical formulas for the friction factor, which can be easily converted into Forchheimer coefficient
using Equation~\eqref{eq:fric_fac}.
Idelchik \cite{idelchik1986handbook} provides several empirical correlations for estimating the friction factor across a perforated plate. 
At finite thickness of the plate and high Reynolds number, Idelchik \cite{idelchik1986handbook} proposes a correlation of the form:

\begin{equation}
    \alpha_y = \dfrac{1}{2\sigma^2 t}\left(     0.5 + 0.24 \sqrt{1-\sigma}(1-\sigma) + (1-\sigma)^2     \right).
    \label{eq:refidelchik}
\end{equation}

Malavasi \textit{et al.} \cite{malavasi2012pressure} suggest an alternative relationship of the form:

\begin{equation}
        \alpha_y = \dfrac{1}{2C^2\sigma^2t}\left( \sqrt{1-\sigma^2 -\sigma^2 C^2} - C\sigma \right)^2,
\end{equation}

where $C$ is a discharge coefficient that depends upon the geometrical parameters of the orifice and the Reynolds number. 
Similarly, Kast \textit{et al.}\cite{kast2010l1} proposes the following relationship:

\begin{equation}
    \alpha_y = \dfrac{1}{2 \sigma^2 t}\left(\left(\dfrac{1}{C} -1\right)^2 + (1-\sigma)^2\right).
    \label{eq:ref2}
\end{equation}

According to Miller ~\cite{miller1978internal} the Forchheimer coefficient can be expressed as:

\begin{equation}
    \alpha_y = \dfrac{C_0(1-C_c\sigma)}{C_c^2\sigma^2t},
\end{equation}

where $C_0$ is a coefficient that depends on $t/d$ and $C_c$ is the jet contraction coefficient. 
Holt \textit{et al.} \cite{holt2011cavitation} present a piecewise function for the Forchheimer coefficient,

\[ \alpha_y=  \left\{
\begin{array}{ll}
      \dfrac{1}{2t}\left(2.9-3.79\dfrac{t}{d}\sigma^{0.2}+1.79\left(\dfrac{t}{d}\right)^2\sigma^{0.4}\right)K &  \quad \dfrac{t}{d}\sigma^{0.2} < 0.9\\
      \dfrac{1}{2t}\left(0.876-0.069\dfrac{t}{d}\sigma^{0.2}\right)K &  \quad \dfrac{t}{d}\sigma^{0.2} >0.9, \\
\end{array} 
\right. \]

where    $K = 1- 2/\sigma + 2/\sigma^2(1-1/C_C +1/(2C_C^2))$.

\end{appendices}

\bibliography{sn-bibliography}
\end{document}